\documentstyle[epsf,twoside,fleqn,espcrc2]{article}

\newcommand{\AmS}{{\protect\the\textfont2
  A\kern-.1667em\lower.5ex\hbox{M}\kern-.125emS}}

\def\simm#1{\mathop{\vtop{\ialign{##\crcr
        $\hfil\displaystyle{#1}\hfil$\crcr\noalign{\kern0.5pt\nointerlineskip}
        $\sim$\crcr\noalign{\kern0.5pt}}}}\limits}

\def\be{\begin{equation}}
\def\ee{\end{equation}}
\def\ba{\begin{eqnarray}}
\def\ea{\end{eqnarray}}

\title{
\vspace*{-35pt}
{\normalsize \hfill {\sf UTCCP-P-26}} \\
\vspace*{-6pt}
{\normalsize \hfill {\sf Sept.\ 1997}} \\
CP-PACS Result for the Quenched Light Hadron Spectrum%
\thanks{Talk presented by K.\ Kanaya at Lattice97, Edinburgh, Scotland, 
22--26 July 1997.}
}

\author{
CP-PACS Collaboration: 
\vspace{1mm}
S.\ Aoki\rlap,\address{Institute of Physics,
University of Tsukuba, Tsukuba, Ibaraki 305, Japan}
G.\ Boyd\rlap,\address{Center for Computational Physics, 
University of Tsukuba, Tsukuba, Ibaraki 305, Japan}
R.\ Burkhalter\rlap,$^{\rm b}$
S.\ Hashimoto\rlap,\address{Computing Research Center, 
High Energy Accelerator Research Organization (KEK), Tsukuba, Ibaraki 305, 
Japan}
N.\ Ishizuka\rlap,$^{\rm a,b}$
Y.\ Iwasaki\rlap,$^{\rm a,b}$
K.\ Kanaya\rlap,$^{\rm a,b}$
Y.\ Kuramashi\rlap,\address{Institute of Particle and Nuclear Studies,
High Energy Accelerator Research Organization (KEK), Tsukuba, Ibaraki 305, 
Japan}
M.\ Okawa\rlap,$^{\rm d}$
A.\ Ukawa\rlap,$^{\rm a}$
and
T.\ Yoshi\'e$^{\rm a,b}$
}

\begin{document}

\begin{abstract}

The quenched hadron spectrum in the continuum obtained 
with the Wilson quark action in recent simulations on the CP-PACS 
is presented.  Results for the light quark masses and the QCD scale 
parameter are reported.

\end{abstract}

\maketitle

\section{Introduction}

Ever since the pioneering work of Hamber and 
Parisi\cite{Hamber_Parisi_81} and of Weingarten\cite{Weingarten_82}, 
the calculation of the light hadron spectrum has been 
regarded as a fundamental subject in lattice QCD.  
The extensive simulations on the GF11 computer\cite{GF11} have 
established that the quenched spectrum agrees with experiment 
within 5--10\%. 
The next, and final, step within quenched QCD would be 
to further advance the calculational 
accuracy so that possible deviations from the experimental spectrum 
are made manifest.  In this article we report the main results 
recently obtained with the CP-PACS computer towards this goal.

\section{Simulation}

Our calculation is made with the plaquette gluon action and 
the Wilson quark action.
We employ a spatial size of $L_sa\approx 3$~fm to avoid finite-size effects, 
and $a^{-1}\simm{>} 2$~GeV to facilitate the continuum extrapolation.  
These requirements lead to the selection of $\beta$ and lattice sizes
listed in Table~\ref{tab:parameter}.  Gauge configurations are 
generated with the over-relaxation and heat bath algorithms mixed 
in a 4:1 ratio. Hadron propagators are calculated 
for the five quark masses corresponding to 
$m_\pi/m_\rho\approx 0.75$, 0.7, 0.6, 0.5 and 0.4, the last point 
being a step closer to the chiral limit than hitherto attempted.
All unequal quark mass combinations allowed for degenerate $u$ and $d$ 
quarks are taken.
We smear the quark source with the function 
$\Psi(r) = A \, \exp[-B(m_\rho,m_q) \, r]$, with $B\approx 0.33$~fm.
Hadron masses are extracted by uncorrelated fits together with a single 
elimination jackknife procedure for estimating errors.  

\begin{table}[t]
\setlength{\tabcolsep}{0.3pc}
\caption{Parameters of simulation.}
\vspace{-1mm}
\begin{center}
\begin{tabular}{lcccc}
\hline
$\beta$ & lattice & $a^{-1}$[GeV] & $L_s a$[fm]& $N_{\rm conf}$ \\
\hline
5.90 & $32^3\times\,56$  & 1.97(1) & 3.21(2) & 800 \\
6.10 & $40^3\times\,70$  & 2.60(1) & 3.04(2) & 600 \\
6.25 & $48^3\times\,84$  & 3.12(2) & 3.03(2) & 420 \\
6.47 & $64^3\times 112$  & 4.16(4) & 3.03(3) &  91 \\
\hline
\end{tabular}
\label{tab:parameter}
\end{center}
\vspace{-11mm}
\end{table}

\begin{figure}[t]
\vspace{-9mm}
\begin{center} \leavevmode
\epsfxsize=7.0cm \epsfbox{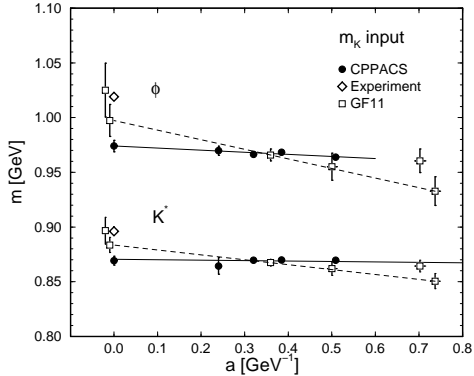}
\end{center}
\vspace{-18mm}
\caption{Continuum extrapolation of $K^*$ and $\phi$ masses 
with the $m_K$-input.
GF11 estimate for infinite volume\protect\cite{GF11} is plotted left 
at $\!a\!=0$.}
\label{fig:mesonE}
\end{figure}

\begin{figure}[t]
\vspace*{-13mm}
\begin{center} \leavevmode
\epsfxsize=7.0cm \epsfbox{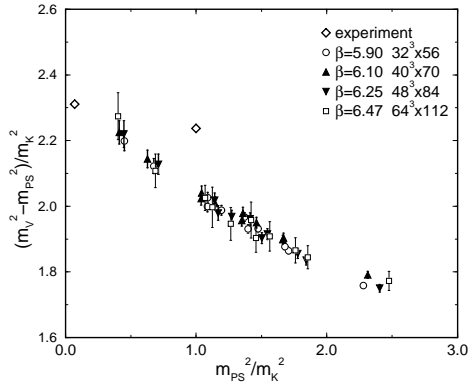}
\end{center}
\vspace{-16mm}
\caption{Meson hyperfine splitting.}
\label{fig:hyperfine}
\vspace{-7mm}
\end{figure}

\section{Meson spectrum}

We find $m_{PS}^2$ for pseudoscalar and $m_V$ for vector mesons to be 
well described by a linear function of $1/K$, which we adopt for 
our chiral fits.  The physical point for the 
degenerate $u$ and $d$ quarks is fixed by 
$m_\pi$(135) and $m_\rho$(769), and we use $m_K$(498) or
$m_{\phi}$(1019) for the $s$ quark.

In Fig.~\ref{fig:mesonE} we show the continuum extrapolation of strange 
vector meson masses with the $m_K$ as input. A linear fit in $a$ yields the 
continuum value which is smaller than the experimental value 
by 3\% for $K^*$ and by 4\% for $\phi$, to be compared with the 
statistical error of 0.5\%. 
If we take $m_\phi$ as the input for the $s$ quark, 
$m_{K^*}$ agrees with experiment to 0.6\%, 
while $m_K$ is higher by 9\%.     

The origin of the discrepancy can be traced to the hyperfine splitting 
shown in Fig.~\ref{fig:hyperfine}.  Our results, which scale well,  
decrease too fast as meson mass increases.
The slope $b$ of the curve is related to the $J$ parameter 
through $J=(1+b)/2$.  A fast 
decrease translates into a small value of $J$; we obtain 
$J=0.377(11)$ in the continuum limit.

\begin{figure}[t]
\vspace{-9mm}
\begin{center} \leavevmode
\epsfxsize=7.0cm \epsfbox{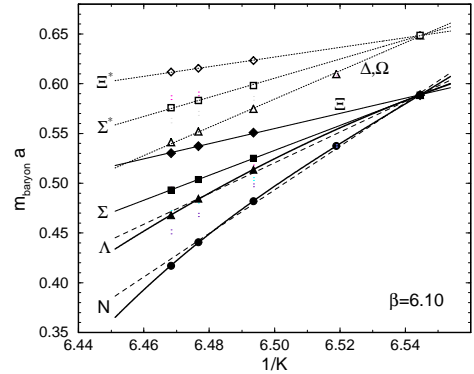}
\end{center}
\vspace{-18mm}
\caption{Baryon masses as a function of $1/K$ for the light quark 
at $\beta=6.1$. For non-degenerate baryons, results with 
the heavier $s$ quark ($m_\pi/m_\rho\!\approx\! 0.75$) are plotted.}
\label{fig:baryonC}
\vspace{-6mm}
\end{figure}

\begin{figure}[t]
\vspace{-9mm}
\begin{center} \leavevmode
\epsfxsize=7.0cm \epsfbox{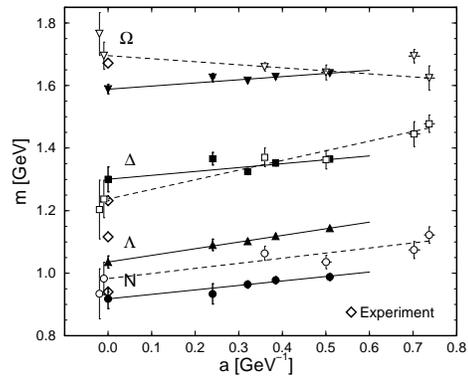}
\end{center}
\vspace{-18mm}
\caption{Continuum extrapolation of baryon masses. Open symbols show 
GF11 results.}
\label{fig:baryonE}
\vspace{-6mm}
\end{figure}

\section{Baryon spectrum}

We show our typical chiral extrapolation for baryon masses in 
Fig.~\ref{fig:baryonC}. We find all baryon masses to be linear in 
$1/K$ except for $m_N$ and $m_\Lambda$,  which show 
a clear negative curvature. We choose a fit cubic in $1/K$ for  
$m_N$, quadratic for $m_\Lambda$, and linear for the other baryon masses.

The negative curvature significantly lowers the nucleon mass compared to 
previous results at a finite lattice spacing, as shown in 
Fig.~\ref{fig:baryonE}.  
The continuum value obtained from a linear extrapolation is smaller than
experiment by 2.3\%, albeit consistent within the statistical error of 3\%.
The mass of $\Delta$ is 6\% higher, with a statistical 
significance of 1.7 standard deviations.

The nucleon mass at each $\beta$ depends on the form adopted 
for the chiral extrapolation.  Including the form from quenched chiral 
perturbation theory, the continuum values,  
however, agree within 3\%. 

The extraction of the continuum value is most uncertain 
for the nucleon, $\Delta$, and $\Sigma^{*}$ masses, for which 
the errors in the continuum limit are about 3\%.  
For all the other channels a linear chiral extrapolation 
(quadratic for $m_\Lambda$) 
followed by a linear continuum extrapolation fits our data very well, 
resulting in errors of 1--2\%.

\begin{figure}[t]
\vspace{-9mm}
\begin{center} \leavevmode
\epsfxsize=7.5cm \epsfbox{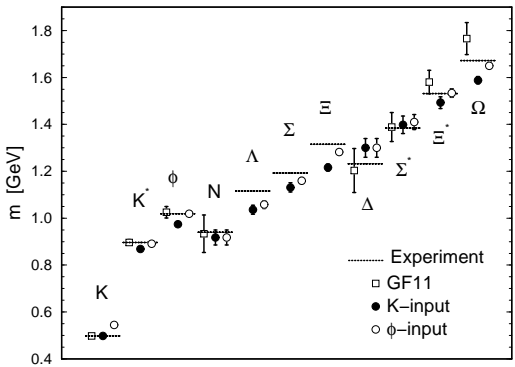}
\end{center}
\vspace{-20mm}
\caption{Quenched hadron spectrum in the continuum limit.}
\label{fig:spectrum}
\vspace{-6mm}
\end{figure}

Our final result for the baryon spectrum is shown 
in Fig.~\ref{fig:spectrum}, together with the meson spectrum for 
completeness.

In the octet the strange baryon masses are lower than experiment, 
by 5--8\% using $m_K$ as input, and by 3--5\% for $m_\phi$ as input.  
The Gell-Mann-Okubo (GMO) relation, however, is well satisfied, at 1\% 
for both cases.

For the decuplet the GMO relation takes the form of 
an equal spacing rule. This is also well satisfied with our results, 
the three spacings mutually agreeing within 5--10\%.  
The average spacings, however, are 
too small by 30\% with $m_K$ as input and by 20\% for $m_\phi$ as input.

The agreement of the baryon spectrum with experiment is generally better
if we use $m_\phi$ as input, particularly for the decuplet.  
However, we have to remember that 
the $K$ meson mass differs from experiment by 9\% in this 
case, which is a 7 standard deviation effect.

\section{Light quark masses}

\begin{figure}[t]
\vspace{-9mm}
\begin{center} \leavevmode
\epsfxsize=7.2cm \epsfbox{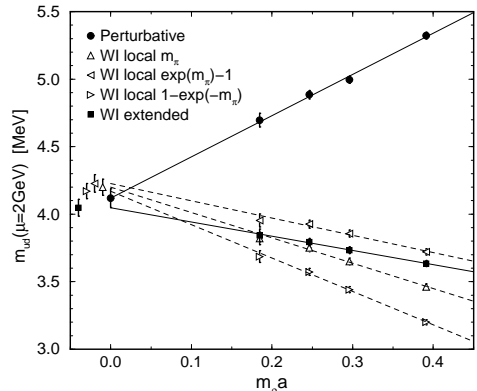}
\end{center}
\vspace{-17mm}
\caption{$\overline{m}\!=\!(m_u+m_d)/2$ at 2~GeV in the 
${\overline{\rm MS}}$ scheme calculated by various definitions.
}
\label{fig:Mud}
\vspace{-6mm}
\end{figure}

We calculate light quark masses with 
the perturbative definition $m_qa=Z_m (1/K-1/K_c)/2$ and with those 
based on the axial Ward identity $\nabla_\mu\!A_\mu\!=\!2m_qaP$.
The two methods give results considerably different at finite 
lattice spacings.  However, extrapolated linearly in the lattice 
spacing, they converge to a common value within 5\% as shown in 
Fig.~\ref{fig:Mud}. 
For average $u$,$d$ quark mass at 2~GeV in the 
$\overline{\rm MS}$ scheme,
we obtain $\overline{m}\!=\!4.1(2)$~MeV.  
For the $s$ quark mass, we find $m_s\!=\! 135(7)$MeV and 111(4)MeV 
using $m_\phi$ and $m_K$. 
The discrepancy reflects the quenching error 
in the strange meson spectrum. 

\section{QCD scale parameter}

We estimate $\Lambda_{\overline{\rm MS}}$ using $\alpha_P(3.40/a)$ and 
$\alpha_{\overline{\rm MS}}(\pi/a)$ including the 3-loop correction in 
the $\Lambda$ parameter and the scale determined from $m_\rho$. 
The couplings are estimated from the average plaquette
with 2-loop perturbation theory.  A linear continuum extrapolation 
of the two results converge and yields 
$\Lambda_{\overline{\rm MS}}\!=\!230(5)$~MeV 
where the error includes extrapolation uncertainties.

\vspace{2mm}

This work is supported in part by the Grants-in-Aid
of Ministry of Education
(Nos.\ 08NP0101,~08640349,~08640350,~08640404,~092462-06, 09304029, 09740226). 
GB and RB are supported by JSPS.

\end{document}